\documentstyle[aps,preprint]{revtex}

\begin{document}

\draft

\title{
Spin-twist driven persistent current in a strongly correlated \\
two-dimensional electron system:\\ a manifestation of the gauge field
}

\author{K. Kusakabe}
\address{
Institute for Solid State Physics, University of Tokyo, Roppongi,
Tokyo 106, Japan
}

\author{H. Aoki}
\address{
Department of Physics, University of Tokyo, Hongo, Tokyo 113, Japan
}

\maketitle

\begin{abstract}
\begin{normalsize}
A persistent current, coupled with the spin state,
of purely many-body origin is shown to exist
in Nagaoka's ferromagnetic state in two dimensions (2D).
This we regard as a manifestation of a gauge field,
which comes from the surrounding spin configuration and
acts on the hole motion, being coupled to the Aharonov-Bohm flux.
This provides an example where the electron-electron interaction exerts
a profound effect involving the spins
in clean two-dimensional lattice systems
in sharp contrast to continuum or spinless fermion systems.
\end{normalsize}
\end{abstract}

\narrowtext

Persistent current is a highlight of the physics of electrons
on mesoscopic scales.
One major interest is how the electron-electron interaction
affects the persistent current.
The issue was originally posed to answer whether the interaction
can offset the reduction of the current due to disorder.
The many-body effect has been examined for both continuum systems and
lattice systems with various geometries such as a one-dimensional ring
or a
torus\cite{ambegaokar,losslu,smith,abraham,weiden,berkovits,bouzerar,yoshioka,avishai}.

In clean {\it continuum} systems, however, the electron-electron
interaction, by itself, exerts {\it null} effect on the persistent current,
which is exactly shown in one dimension (1D)
and expected to persist in two dimensions (2D) and higher ones,
unless the interaction creates low-energy excitations
below the lowest excitation of the noninteracting system\cite{weiden}.

The {\it lattice} system is a different story, even when clean,
for two reasons.
First, a lattice system can undergo
Mott's metal-insulator transition at some
band filling or at some critical interaction strength,
so that the interaction may suppress the persistent current around
the transition\cite{shastry,abraham}.

However, a truly interesting effect of interaction in lattice systems
should appear when one considers the spin degrees of freedom.
An effect of spin is known to appear as a
change in the Aharonov-Bohm (AB) period (in units of the flux quantum,
$\Phi_0=hc/e$) and amplitude, originally shown for the noninteracting model
by averaging over the number of electrons\cite{loss} and
later found in interacting systems
such as the one-dimensional Hubbard model with a fixed number
of electrons in the dilute and/or strong-coupling limit.\cite{kusmartsev}
Change in the orbital magnetism and current is also found
in the finite-$U$ Hubbard model
with the spin-orbit interaction\cite{fujimoto}.

In the present paper we show that
an interplay between charge and spin degrees of freedom
can result in a qualitatively different phenomenon, i.e., a persistent
current ``driven by'' a twist in the spin structure with the external flux
in a strong-$U$ Hubbard model in 2D.

In strongly correlated systems, the ground state is often sensitive to
the boundary condition.
For instance, Nagaoka's ferromagnetic ground state\cite{nagaoka},
which is an extreme manifestation of the correlation for large $U$
in the immediate (one-hole doped) vicinity of the Mott insulator,
evolves into a spin-singlet state when
the antiperiodic boundary condition is adopted in one direction.\cite{riera}
We can then expect that the boundary-condition dependence
will signify a large response to the AB flux and a
persistent current via the Byers-Yang theorem.\cite{byers}
This does indeed turn out to be the case.
We shall trace this back to a gauge potential inherent in
lattice systems, where the gauge potential has an ability to
cope with the frustration caused by the twisted boundary condition,
a situation distinct from continuum systems.

We consider the two-dimensional Hubbard model on a torus to get rid of edges,
and pierce an AB flux $\Phi$ through the opening.
This amounts to a frustration, or a change
in the boundary condition along $y$ for the wave function $\phi$ to
\begin{eqnarray}
\phi (\ldots,{\bf x}_i+N_y {\bf u}_y,\ldots)&=&\exp \,
({\rm i}2\pi\Phi/\Phi_0) \phi (\ldots,{\bf x}_i,\ldots) \; , \nonumber
\end{eqnarray}
where ${\bf x}_i$ is the position of the $i$th electron,
and ${\bf u}_a$ ($N_a$) the unit vector
(the number of sites) in the $a$ $(=x,y)$ direction with
the lattice constant being unity.

We have numerically obtained the change in the energy spectrum with $\Phi$
at $U=\infty$ for 15 (14) electrons on a $4\times 4$-site system,
i.e., one (two) -hole doped from half filling) in Fig. \ref{grsp16}.
We immediately recognize that there is a dramatic series of level crossings
for low-lying states in such a way that
the total spin of the ground state $S$ oscillates between
Nagaoka's {\it full} polarization
(which occurs at $\Phi =0$ for one hole or at
the half-flux for two holes)
and vanishing $(S\simeq 0)$ polarization.

This oscillation gives rise to a persistent current
$I$ (Fig. \ref{curr16}) via the Byers-Yang theorem,\cite{byers}
\begin{equation}
I=- c \frac{\partial \langle \phi | H(\Phi )
|\phi \rangle }{\partial \Phi} \; ,
\end{equation}
where $|\phi \rangle $ is assumed to be normalized.
Thus, not only the magnitude but the overall profile of the current
deviate from the spinless electron system (the curve for the fully
polarized state).

The flux-induced oscillation in $S$ found here
is totally incompatible with the ordinary Stoner ferromagnetism.
In the Stoner picture,
the lowest charge excitation with a spin flip is the Stoner excitation, for
which a Stoner gap opens over the whole Brillouin zone
for a fully polarized state,
so that the Stoner excitation cannot contribute to the persistent current.
The collective spin wave also remains in an excited state
when $\Phi$ is varied.
For partially polarized states, the Stoner excitation continuum becomes
gapless and a finite $\Phi$ can induce single spin flips, but this will only
result in a change of $S$ by unity.
In Nagaoka's state, by contrast, $S$ oscillates between the largest and
smallest possible
values, which is due to an anomaly in the gap and spin stiffness where
the spin mass vanishes like the inverse system size\cite{nagaoka,bry,ka1}.

In a doped Mott insulator, which we have at hand,
a fundamental deviation from the spinless fermions arises
from an interplay of the charge (hole) motion and
the surrounding spin configuration, which is indeed responsible
for anomalous excitations.
To show this we can exploit the formulation of Ref. \cite{nagins},
where we write down an eigenfunction of the one-hole system in a
``Bloch's form'' as
\begin{equation}
|\phi \rangle = \sum_{{\bf x}_{\rm h}} \exp \,
({\rm i}{\bf k} \cdot {\bf x}_{\rm h}) \sum_{\{ \sigma_j \} }
f(\sigma_1 , \ldots , \sigma_{N_{\rm e}} ) \: c^\dagger_{{\bf x}_1,\sigma_1}
\cdots c^\dagger_{{\bf x}_{N_{\rm e}},\sigma_{N_{\rm e}}} | 0 \rangle .
\label{bfunc}
\end{equation}
Here ${\bf x}_{\rm h}$ is the position of the hole,
$c^\dagger_{{\bf x}_j,\sigma_j}$
creates an electron with spin $\sigma_j$ at ${\bf x}_j$,
and $N_{\rm e}$ $(=N-1)$ is the number of electrons
in an $N$-site system.
The advantage of working with a spin wavefunction
$f(\sigma_1 , \ldots , \sigma_{N_{\rm e}} )$
around the hole is the following.
If the order $({\bf x}_1,\ldots ,{\bf x}_{N_{\rm e}})$
according to which the creation operators are multiplied, is specified by the
coordinate {\it relative to} ${\bf x}_{\rm h}$,
$f$ no longer contains ${\bf x}_{\rm h}$,
and the equation determining the spin configuration around the hole
is simplified:
A hole hopping from ${\bf x}_{\rm h}$ to ${\bf x}_{\rm h} + {\bf u}_a$
is mapped onto a spin operation that is
almost a uniform translation of the spins in the inverse direction,
except for the spin at ${\bf x}_{\rm h} + {\bf u}_a$,
which has to move to ${\bf x}_{\rm h} -{\bf u}_a$.
If we write the whole operation as ${\cal {T}}_a$,
we end up with a spin Hamiltonian,
\begin{equation}
{\cal H}_{\rm spin} =
   - t \sum^d_{a=1} \{ \exp \, (i{\bf k} \cdot {\bf u}_a ) \: {\cal T}_a
+ \exp \, (-i{\bf k} \cdot {\bf u}_a ) \: {\cal T}^{-1}_a \} \; .
\end{equation}
In this formalism, $\Phi$ simply shifts the $k$ points uniformly
as ${\bf k} \rightarrow {\bf k}+(0,2\pi\Phi /(N_y \Phi_0))$.

We can then identify the origin of the many-body effect on the current as the
gauge potential in the following sense.
When we decompose the motion of electrons
into the center-of-mass and internal ones,
$\Phi$ only couples to the former, which is completely decoupled
from the latter for continuum systems.
Thus, a persistent current cannot deviate from the free-fermion
behavior\cite{weiden}.
Even if the interaction creates low-energy excitations leading to
level crossings for finite $\Phi$, the
curvature of the spectral flow $[E_i(\Phi)]$ does not deviate
from that for the free fermions.

In sharp contrast, the center-of-mass motion does disturb
the relative coordinates on a lattice.
Namely, a shift in ${\bf x}_{\rm h}$,
which coincides with the center-of-mass here,
affects the relative coordinates between electrons through
the electron correlation (i.e., excluded double occupancies).
In the present formalism, the coupling of the two coordinates
appears as a discrepancy of ${\cal T}_a$ from
the uniform translation.\cite{1dcurr}
The spin wave function is thus affected by the flux-coupled charge motion,
and changes its structure with $\Phi$
in two- (or higher) dimensional lattice systems.

Here we notice that the spin configuration
does not flow into an antiferromagnetic (AF) state with $\Phi$,
since this would impede the motion of the hole.
In fact, the state
that takes over Nagaoka's state at $\Phi_0 /2$ is
a {\it twisted} spin structure,
or a spin-density wave whose wavelength is as large as the sample size.
The spiral structure is confirmed here from the spin-spin correlation
for a $(\sqrt{26} \times \sqrt{26})$-site system displayed
as an inset of Fig. \ref{grsp16}(a), where
the spin polarization is seen to rotate gradually along
the $y$ direction in which the AB flux exerts a twist.

Curiously, it is exactly this spiral state that is the first excited state
from Nagaoka's at $\Phi=0$.
Among the lowest-spin ($S=1/2$) states, this (transfer-stabilized)
spiral-spin state has a lower energy than the (exchange-stabilized)
AF state until a level crossing occurs at a critical value of
$U$ $[(=167t$, i.e., $J\equiv 4t^2/U=0.024t$
for a $(\sqrt{26} \times \sqrt{26})$-site system)$]$
as $U$ is decreased.

The same occurs for a two-hole system, where the ground state starts off
with the spiral spin state
at $\Phi =0$ (for small enough $J$).\cite{nagins}
The state is characterized by a spin-spin correlation $S({\bf Q})$
having four peaks at the wave number ${\bf Q} = (\pm 2\pi/N_x, \pm 2\pi/N_y)$
$[$inset of Fig. \ref{curr16} (b)$]$. The $S=0$ state
changes into a two-peaked one
continuously with $\Phi$.

Such twisted spin structures have in fact been analyzed
to discuss the stability of Nagaoka's state.  Dou\c{c}ot and Wen have
introduced a twisted spin state,\cite{dw}
where they assume that a hole hops on a frozen spin background.
There, they have shown that a properly chosen spin texture can remove
some of the frustration caused by the external flux.
Although the Dou\c{c}ot-Wen function is an approximate trial one,
which is not even an eigenstate of the total spin,
the function, when cast into our Bloch form,
does serve a heuristic purpose of identifying two kinds of Bloch momenta.
We can first call ${\bf k}$, which is coupled
to ${\bf x}_{\rm h}$ in Eq. (\ref{bfunc}),
the charge Bloch wave number.
The Dou\c{c}ot-Wen spiral configuration is frozen in the rest frame,
but is now specified by the position of its loop
${\bf r}_0=(x_0,y_0)$ (where the spin wave function has a phase
slip of $\pi$) in our frame moving with the hole.
We can take a
linear combination of spin configurations having different ${\bf r}_0$
with a spin Bloch wave number ${\bf q}$.
Let $\theta = 2\pi/N_y$ be the wave number
of the spin twist in the Dou\c{c}ot-Wen function.
The variational function then reads
\begin{equation}
\label{dwfunc}
|\varphi \rangle = \sum_{{\bf x}_{\rm h}} \exp \, ({\rm i}{\bf k}
\cdot {\bf x}_{\rm h})
\frac{1}{N} \sum_{x_0,y_0} e^{i{\bf q \cdot r}_0}
\prod_{\stackrel{(x,y)}{\neq {\bf x}_{\rm h}}} \otimes \Biggl|
\begin{array}{c}
\cos \, [\frac{\theta}{2} (y+y_0)] \\
\sin \, [\frac{\theta}{2} (y+y_0)]
\end{array}
\Biggr\rangle_{(x,y)} \; ,
\end{equation}
where $|\mbox{}^\alpha_\beta\rangle_{(x,y)}$ is the spinor at site $(x,y)$
and we have again assumed that the order in the product $\prod_{(x,y)}$
is according to the rule described below Eq. (\ref{bfunc}).
If we choose ${\bf q} =(0,\theta /2)$ for the spin Bloch wave number,
the variational energy becomes
\begin{equation}
\varepsilon = -2t\cos \, k_x -2t \cos \, \frac{\theta}{2} \cos \, \left( k_y +
\frac{2\pi \Phi}{N_y \Phi_0} - \frac{\theta}{2} \right) \; ,
\label{dwene}
\end{equation}
for $U=\infty$, and is minimized at
the half-flux, where the difference between the
AB phase $2\pi \Phi/N_y \Phi_0$ and the spin phase $\theta/2$
vanishes.

Thus we are led to a picture in which the extra phase
$\theta /2$ arising from the gauge potential (an internal frustration)
from the spin configuration acts to {\it cancel} the
enhancement in the kinetic energy caused by the external (flux-induced)
frustration.
There, the low-spin states becomes advantageous
for the maximal (half-flux) frustration, since they have the
maximum degrees of freedom in the spin space that
enable them to absorb the frustration.

We expect this should generally apply to strongly correlated systems
with restricted motion of charges (i.e., excluded double occupancies),
so that the variation of the energy, and thus the persistent current,
for spinfull electrons should be smaller than those for (the
same number of) spinless electrons.

To be more precise, the spin configuration is disturbed
by the hole motion as stressed\cite{comment}, and
a factor neglected by the Dou\c{c}ot-Wen wave function
is the quantum fluctuation.  This is numerically seen in
a reduction in the absolute value of the nearest-neighbor
spin-spin correlation
from that for the spiral configuration of classical $S=1/2$ spin
in the inset of Fig. \ref{grsp16}(a).
However, the inseparability of the center-of-mass and internal coordinates
and their interference does in general remain
when the configurational
subspace classified by
the total orbital angular momentum and the spin are
`irreducible' (i.e., when any two states within
the subspace are connected by a series of
nonzero matrix elements of the transfer).
Thus, in addition to the well-known
enhancement in the effective mass of a hole, this gives rise to
the spin-modulated persistent current
in clean two-dimensional lattices, unlike two-dimensional continuum systems.

In other words,
a current deviating from the free-electron behavior
should be generic to any lattice systems, such as the $t$-$J$ model,
that have a constrained electron motion due to strong interactions.
For example, in the AF phase in the finite-$U$ Hubbard model or $t$-$J$ model
in 2D,
$E_i(\Phi)$ deviates from the free-electron result as shown by
Poilblanc\cite{poilblanc}.
A difference there from the present situation is that the level crossing
is shown to occur
among the exchange-stabilized $S_{tot}=0$ states (i.e., within the AF phase).
There, we also observe a reduction in the deviation
of the ground-state energy, which may again be explained
from the flux-induced enhancement of the kinetic energy
as relaxed by the gauge field arising from the spin wave function.

Obviously, another important question is the effect of disorder
or the edge effect for open geometries to make the model approach
to actual mesoscopic systems. As for the ferromagnetism,
whether there is a finite ferromagnetic region around
Nagaoka's limit is still an open question.
A finite region for the partial polarization is proposed
from a $t$-$J$ study\cite{putikka}, but
a possibility of an abrupt transition
to an AF state is also suggested\cite{nagins}.
However, we can provide a way to relax the singular situation of Nagaoka's
by going over to two-band Hubbard models, where the {\it double-exchange}
mechanism induces a more stable ferromagnetism
that we can show has close similarities with Nagaoka's\cite{unpub}.
Another prominent example of the stabilized ferromagnetism is the
Hubbard model on flat-band systems\cite{flat}.  There the flat band,
which is fully polarized when half-filled, becomes
metallic when doped, and it will be interesting to
see how the persistent current behaves in
such an itinerant magnet.

The authors would like to thank Professor Yshai Avishai for discussions.
Numerical calculations were performed on HITAC S3800 in the Computer Center,
University of Tokyo.
This work was supported, in part, by a Grant-in-Aid from
the Ministry of Education, Science and Culture, Japan.

\begin{figure}
\caption{
\label{grsp16}
(a) AB flux ($\Phi$) dependence of the lowest state for each of
various total spins in a one-hole doped $(4\times 4)$-site system
with $U=\infty$, in which
the ground state evolves from a ferromagnetic state at $\Phi =0$
down to a spiral-spin state at $\Phi = \Phi_0/2$.
(b) The same for a two-hole doped $(4\times 4)$-site system.
The insets depict the spin-spin correlation function
$\langle S({\bf 0})S({\bf R}) \rangle$
for the spiral-spin ground state with the half-flux in a one-hole doped
$(\protect \sqrt{26} \times \protect \sqrt{26})$-site system
$[$the arrow in (a)$]$ and
the spin structure factor $S({\bf Q})$ for the ground state
without flux in a two-hole doped
$(\protect \sqrt{20} \times \protect \sqrt{20})$-site system
$[$the arrow in (b)$]$ at $U=\infty$ ($J=0$).
}
\end{figure}

\begin{figure}
\caption{
\label{curr16}
The persistent current derived from the previous figure via
the Byers-Yang theorem for the $(4\times 4)$-site Hubbard model
for one hole (a) or two holes (b)
with $U=\infty$.
}
\end{figure}


\begin{references}
\bibitem{1dcurr}In one dimension, ${\cal T}_a$ reduces to a rotation
on the ring for one hole, but the spin momentum,
$2n\pi /(N-1)$ ($n$: an integer), is incommensurate with $k$
resulting in an eigenenergy, $-2t \cos \, (k+2\pi\Phi/N\Phi_0 -2n\pi/(N-1))$.
Although we need the Bethe-ansatz solution for general
$N_{\rm e}$ (Ref. \cite{kusmartsev}), $N_{\rm e}$-flux periodicity
is derived for $N_{\rm e}=N-1$ already from the above expression.
\end{references}
\end{document}